\documentclass[pdf,color]{UoBnote}

\usepackage{color}

\usepackage{caption2}

\setcaptionmargin{1cm}

\author{Stefan Hild, Simon Chelkowski and Andreas Freise \\ \\ email:hild@star.sr.bham.ac.uk}

\title{Pushing towards the ET sensitivity using `conventional' technology}
\shorttitle{ET sensitivity study}
\date{\today}
\issue{2}

\let \includegraphics \includegraphics

\begin{document}

\maketitle

\tableofcontents
\vspace{1cm} 
\hrule height 1pt
\vspace{1cm}


\section{The Scope of this Document}

Recently, the design study `Einstein gravitational wave Telescope' (ET) has been funded within the European
FP7 framework~\cite{ET}. The design study represents the first coordinated effort towards the
design of a third-generation gravitational wave detector. The ambitious goal of 
this project is to provide a conceptual design of a detector with a hundred 
times better sensitivity than currently operating detectors, which corresponds to 
a capability of scanning a one million times larger volume of the 
universe! It is expected that this challenging goal requires  
the development and implementation of new technologies, which go beyond the
concepts employed for the first and second detector generations, especially for the reduction
of quantum noise. One task within the
design study is to review such new technologies and study their feasibility
for gravitational wave detection. In this context, also the application of completely 
different detector geometries and topologies is being discussed \cite{trimi}. 

However, it is a very interesting and educational exercise to imagine
a Michelson interferometer in which conventional technologies have been pushed
to - or maybe beyond - their limits to reach the envisaged sensitivity for the Einstein Telescope.
In this document we present a first sketchy analysis of what modifications and
improvements are necessary to go, step-by-step, from 
second generation gravitational wave detectors to the Einstein Telescope. 
We restrict our analysis to a configuration similar to Advanced LIGO 
(Dual-Recycled Michelson interferometer featuring arm cavities) but with 
an increased arm length of 10\,km and we explicitly put a focus on rather 
conventional techniques.
\begin{figure}[htb]
\begin{center}
\includegraphics [scale=0.4]{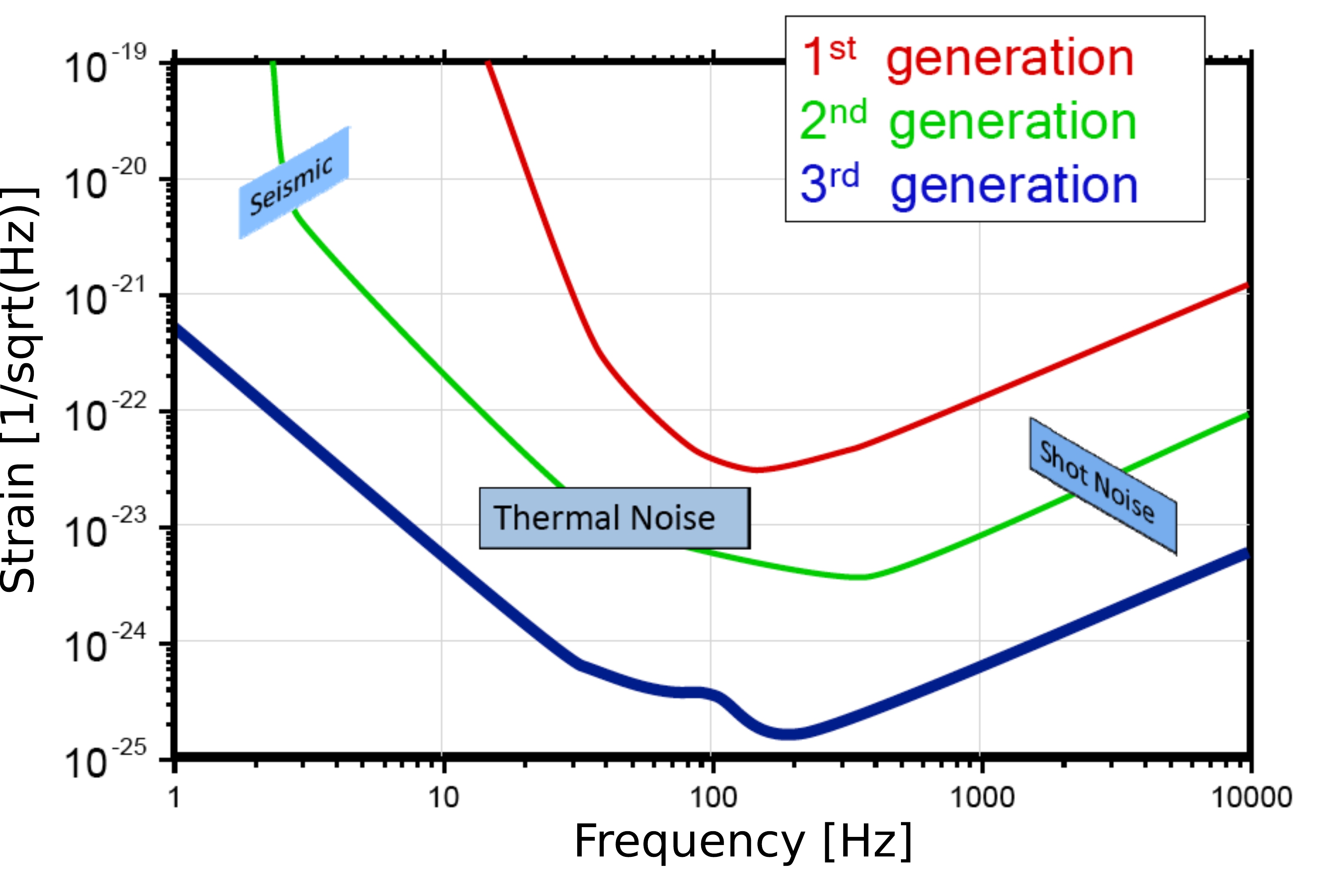}
\end{center}
\caption[]
{Originally proposed target sensitivity of the Einstein Telescope (blue) from \cite{Michele}. 
For comparison a rough estimate of the average sensitivity of first (red) and 
second generation (also called 'advanced') gravitational wave detectors (green) are included.}
\label{fig:target}
\end{figure}

\section{Assumptions and Constraints}
As described above we want to restrict ourselves to 
more or less \emph{conventional} technologies which fulfill at least one of the following two criteria: 
\begin{itemize}
\item The technology was already successfully  demonstrated on prototypes, such as the injection of
squeezed light into a Michelson interferometer or interferometry with cryogenic mirrors.
\item The technology is an up-scaling of currently used technology without
any change of the fundamental physics involved, for instance using suspensions of 50\,m length.
\end{itemize}
Thus, we consider `non-conventional', any technique or technology which
so far only exist on paper, or has been demonstrated only in proof-of-principle
experiments which are very different from the target interferometers. Examples
for such technologies include displacement-noise-free interferometry, optical bars and optical
levers or the use of higher-order Laguerre-Gauss-modes.
These `non-conventional' techniques have been omitted in our analysis.

A further constraint we imposed on our analysis is that we did not consider the possibility
to split ET into separate smaller, optimised interferometers, for example 
a high-power high-frequency device and a low-power low-frequency device. 
Although such an approach might be promising for reducing the quantum noise, it
would not show the limits of conventional techniques as clearly as a single broadband
detector.

Restricting our analysis to a single interferometer also implicates, that we do not
evaluate any properties which are closely related to a potential network of detectors, 
such as for instance sky-coverage, null-stream availability or redundancy.

\section{A GWINC Model for ET}

The Advanced Virgo design group makes use of a well documented noise 
model~\cite{VIR-055A-08}, which is based on the GWINC code \cite{gwinc} 
developed within the LIGO scientific collaboration. We use this
Advanced Virgo noise model as starting point for our investigations. 
\begin{figure}[htb]
\begin{center}
\includegraphics [scale=0.5]{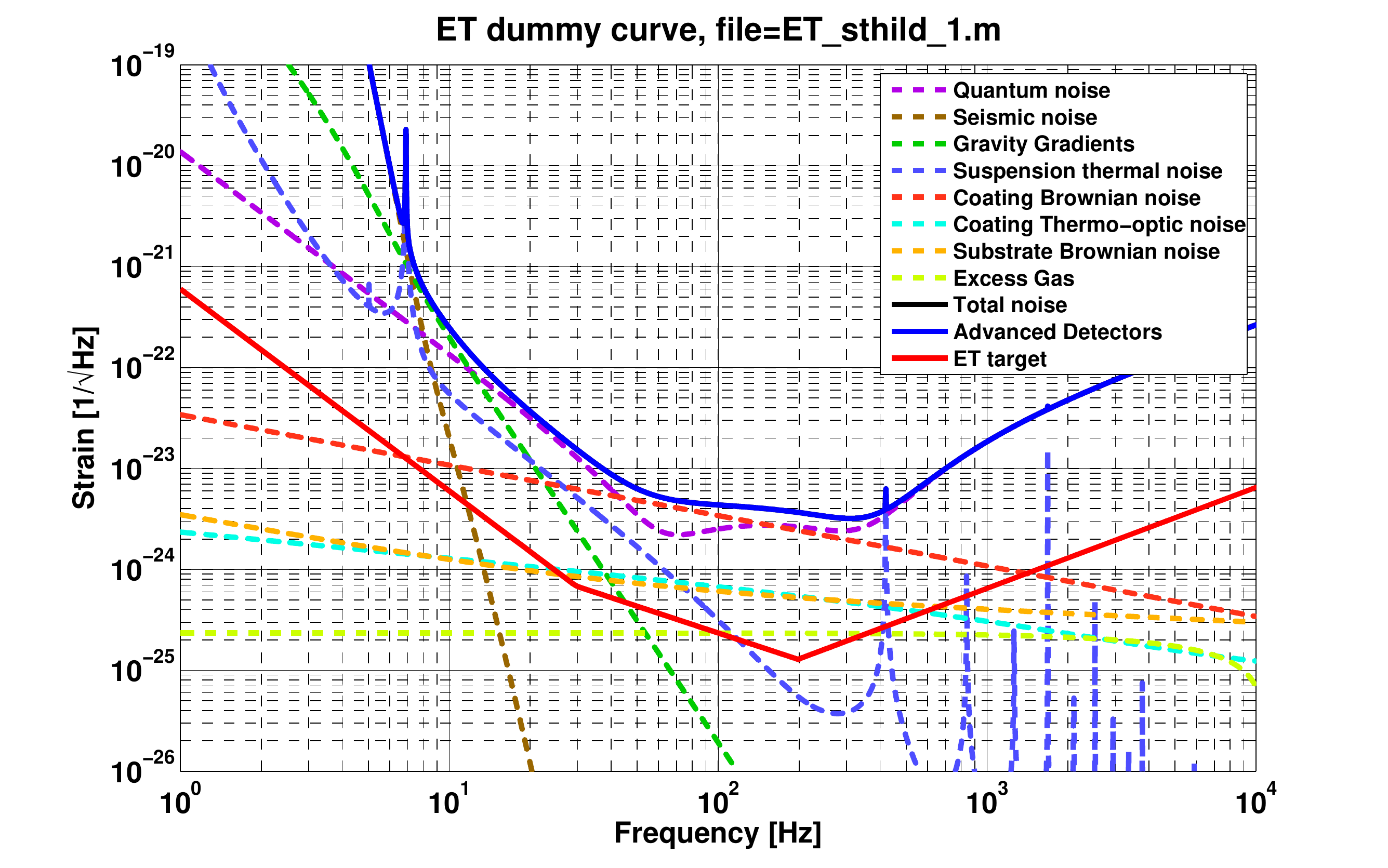}
\end{center}
\caption[]
{Fundamental noise contributions to the sensitivity of a potential 
advanced detector (blue solid line). The solid
red line is an approximation of the ET design target (blue line from Figure \ref{fig:target}).
 Please note that every noise source is at least at some frequencies above the ET design
target, thus we have to improve every single noise contribution to guarantee compatibility
with the targeted ET sensitivity.}
\label{fig:1}
\end{figure}

Figure \ref{fig:1} shows the fundamental noise contributions of an detector with 
basic parameters similar to a possible Advanced Virgo design, using an arbitrary 
detector configuration. Please note that the resulting sensitivity (solid blue line) 
is for an imaginary advanced detector only and is not connected to the ongoing 
design process for Advanced Virgo. The solid red line is an approximation of the 
ET design target (blue line from Figure \ref{fig:target}). Please note that every single
noise source reaches above the ET target, at least for some signal frequencies.
Therefore, each of the displayed noise curves has to be lowered by changing the 
interferometer performance. In the next section we will change step by step 
the relevant parameters of our noise model, starting with a noise model
for an advanced detector, in order to suppress each individual 
noise contribution until we achieve compatibility with the ET design target.

\section{From Advanced Detectors to the ET Design Target}\label{list}

The following list indicates briefly which parameter changes were performed to
go from an advanced detector to the Einstein Telescope. Please
note: The items listed below represent only \textbf{one} possible approach to achieve the
ET target sensitivity. There might be other, perhaps more elegant 
or more feasible scenarios even with conventional technologies. A realistic
detector design will very likely contain some of the techniques labeled as
non-conventional  in this document.

\begin{figure}[htb]
\begin{center}
\includegraphics [scale=0.5]{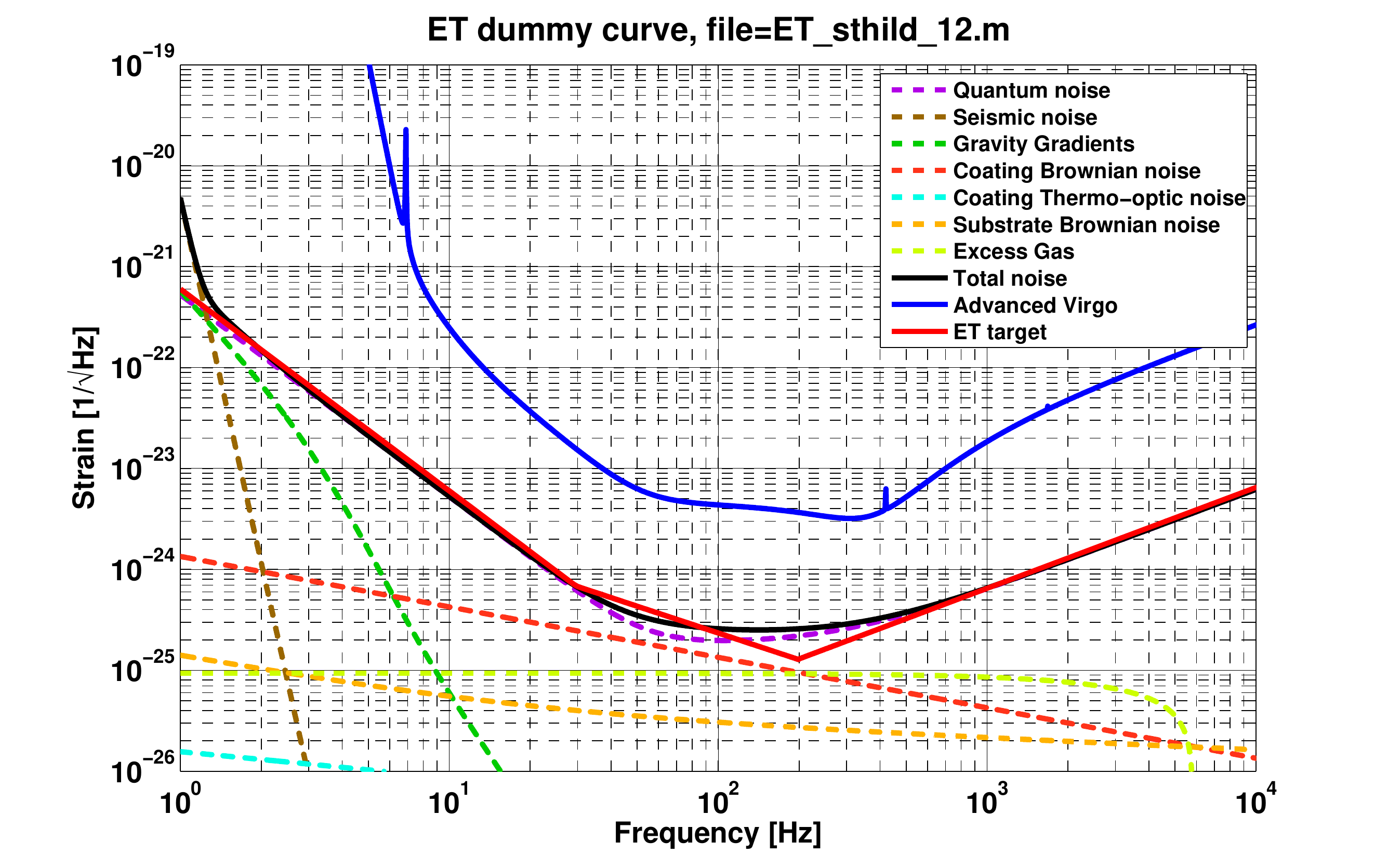}
\end{center}
\caption[]
{Result from the analysis presented in this document. With the changes suggested above 
we can roughly achieve (black solid line) the ET design target (red solid line)
 only using conventional technology.
However, in the end it should be possible to further increase the peak
 sensitivity by simultaneously improving shot noise, coating Brownian noise 
and residual gas pressure. }
\label{fig:12}
\end{figure}

\begin{itemize} 

\item In a first step we \textbf{increased the arm length from 3 to 10\,km} ({\tt ET\_sthild\_2.m}),
which reduces all displacement noises (seismic, gravity gradients, suspension thermal,
coating Brownian, coating thermo-optic and substrate Brownian noise) by a factor 3.3 and the
residual gas pressure noise by about a factor $\sqrt{3.3}$. Please note that all other 
detector parameter, such as the beam size at the test masses are kept constant.

\item Then we adjusted the Signal Recycling from detuned (SR-phase = 0.15) to \textbf{tuned
Signal Recycling} (SR-phase = 0) in order to maximize the detector bandwidth (by 
resonant enhancement of both signal sidebands). 
Additionally  we have modified the Signal Recycling mirror transmittance slightly 
from 11\% to 10\%  as a  compromise of peak sensitivity and detector 
bandwidth ({\tt ET\_sthild\_3.m}). 

\item In a further step we \textbf{increased the laser input power}
 from 125 Watt to 500 Watts
 which yields an intracavity power of about 3\,MW  ({\tt ET\_sthild\_4.m}). The value of 
 500\,Watts has been chosen by eye as a trade off between improved shot noise at high
frequencies and higher radiation pressure noise at low frequencies. 

\item Then we introduced a \textbf{quantum noise suppression factor} of 10\,dB
 ({\tt ET\_sthild\_5.m}), which yields a broadband reduction of photon shot noise and 
radiation pressure noise of about a factor 3.

\item Next we chose to decrease coating Brownian noise by \textbf{increasing
  the beam size} at the main test masses. We enlarged the beam radius (1/$e^2$ in power) 
from 6 to 12\,cm ({\tt ET\_sthild\_6.m}). This requires test mass
radii of curvature of 5070\,m for input and end mirrors.
This increase of beam size reduces coating and substrate thermal noise by about a factor of 2.
In addition, also the residual gas pressure is slightly improved, due to the larger volume of 
the beam. Please note that assuming such large beams has the consequence
that the mirror coatings have to have a diameter of 60 to 70\,cm in order 
to keep the clipping losses within an acceptable range.

\item Since the coating noise was still limiting we also had to consider cooling of the optics.
Consequently we introduced a \textbf{reduction of the temperature} from 290 to 20\,K 
({\tt ET\_sthild\_7.m}). Please note that going to cryogenic temperatures could
involve a change of the test mass material from fused silica to sapphire.


\item The next step was to reduce the seismic noise. We replaced the Virgo super
 attenuator by a \textbf{suspension consisting of 5 stages each 10\,m long}.\footnote{A completely
different but also somewhat conventional approach to tackle the seismic noise would be
to use suspension point interferometry \cite{Matt}.}
We used a very simplistic model for the
suspension transfer function comprising of just 10 simple poles with a corner frequency
of 0.158 Hz. For the seismic we used a simplified spectrum 
($1\cdot 10^{-7}$\,m$/f^2$ for $f>1$\,Hz) of the horizontal seismic 
measured at Virgo site \cite{windmill} ({\tt ET\_sthild\_9.m}). Please 
note that for our new suspensions the suspension thermal noise is considered to be not limiting
the ET sensitivity and has therefore been omitted in the following.

\item In order to further improve the seismic noise we had to go \textbf{underground}. 
We assumed the underground seismic to be about $5\cdot 10^{-9}$\,m$/f^2$ for $f>1$\,Hz which 
corresponds roughly to the seismic level measured in the Japanese Kamioka mine 
(Slide 26 of \cite{Michele}). This is assumed to reduce both the seismic noise as well as the 
gravity gradient noise by a factor of 20 (see {\tt ET\_sthild\_10.m}).

\item Even when going underground the gravity gradient noise seems still to be 
above the ET target sensitivity for frequencies below 6\,Hz (see Figure ET\_sthild\_10
in the Appendix section). Therefore it will  be necessary to find some way to further
reduce this noise by another factor 50 (maybe by a clever shaping of the caves) to 
finally get it below the ET sensitivity target ({\tt ET\_sthild\_11.m}). Please note 
that this somehow \emph{magic} reduction of gravity gradient noise is the only aspect of our 
analysis which is not compatible with our initial definition of \emph{conventional}.

\item Finally, in order to bring the radiation pressure below the ET design target we
had to \textbf{increase the weight of the mirrors} from 42\,kg to 120\,kg
({\tt ET\_sthild\_12.m}).
\end{itemize}
\begin{table}
\begin{center}
\begin{tabular}{|l|c|c|}
\hline 
& advanced detector  & potential ET design \\
\hline 
Arm length & 3\,km & 10\,km \\
SR-phase & detuned (0.15) & tuned (0.0)\\
SR transmittance & 11\,\% & 10\,\% \\
Input power (after IMC) & 125\,W & 500\,W \\
Arm power & ~0.75\,MW & ~3\,MW\\
Quantum noise suppression & none & 10\,dB \\
Beam radius & ~6\,cm & 12\,cm \\
Temperature & 290\,K & 20\,K \\
Suspension & Superattenuator & 5 stages of each 10\,m length\\
Seismic & $1\cdot 10^{-7}\,{\rm m}/f^2$ for $f>1$\,Hz (Cascina) & $5\cdot 10^{-9}\,{\rm m}/f^2$ for $f>1$\,Hz (Kamioka) \\
Gravity gradient reduction & none & factor 50 required (cave shaping) \\
Mirror masses & 42\,kg & 120\,kg \\
BNS range & ~150\,Mpc & ~ 2650\,Mpc\\
BBH range & ~ 800\,Mpc & ~ 17700\,Mpc \\
\hline
\end{tabular}
\caption{Summary of the parameter changes necessary to go from the 
advanced detector sensitivity to the ET design target. \label{tab:summary}}
\end{center}
\end{table}
The final result of our analysis is shown in Figure \ref{fig:12}.
With the changes described above we can achieve 
a sensitivity (black solid line) close to the ET design target (red solid line).
Table~\ref{tab:summary} shows a summary of the parameter changes to transform
an advanced detector into a new detector which achieves a sensitivity
compatible with the ET design target. Our detector model gives an inspiral range for
binary neutron stars of about 2530\,Mpc and of about 17500\,Mpc for binary black hole systems.

\section{Summary and Outlook}
The analysis presented in this document
indicates that it is, in principle, possible to achieve the ET target
sensitivity with \emph{conventional} technology. Only the rather unexplored gravity 
gradient noise needs to be suppressed by non-conventional techniques in order to
be compatible with the ET sensitivity at the very low end of the detection band 
(below 6\,Hz). However, we already know that some of the required up-scalings of existing
technologies will introduce new noise couplings. The best example for that is of
course the high circulating light power which can introduce stability problems to the
arm cavities. Therefore we do not believe that up-scaling existing technologies
is always the best or even a viable option for the design of a third generation
detector. It is important to explore new detector topologies and new technologies 
(such as broadband QND-techniques, Sagnac-topologies, displacement noise free interferometry, 
to list only a few examples) and compare their costs and feasibility to conventional 
technologies. The purpose of this document is simply to provide the parameters shown
in Table~\ref{tab:summary} as a reference for that work.

\section{Acknowledgment} 
The research leading to these results has received funding from the European
 Community’s Seventh Framework Programme (FP7), under the Grant Agreement
 211743, ET project.

\clearpage

\section{Appendix: Intermediate Steps of our Analysis}

This section contains all noise budgets from each intermediate step of our analysis presented
 in Section \ref{list}. The file names in the title of each plot corresponds
to the item containing the same file name as reference.

\vspace{2cm}
\begin{center}
\includegraphics [viewport=30 0 710 480,scale=0.47]{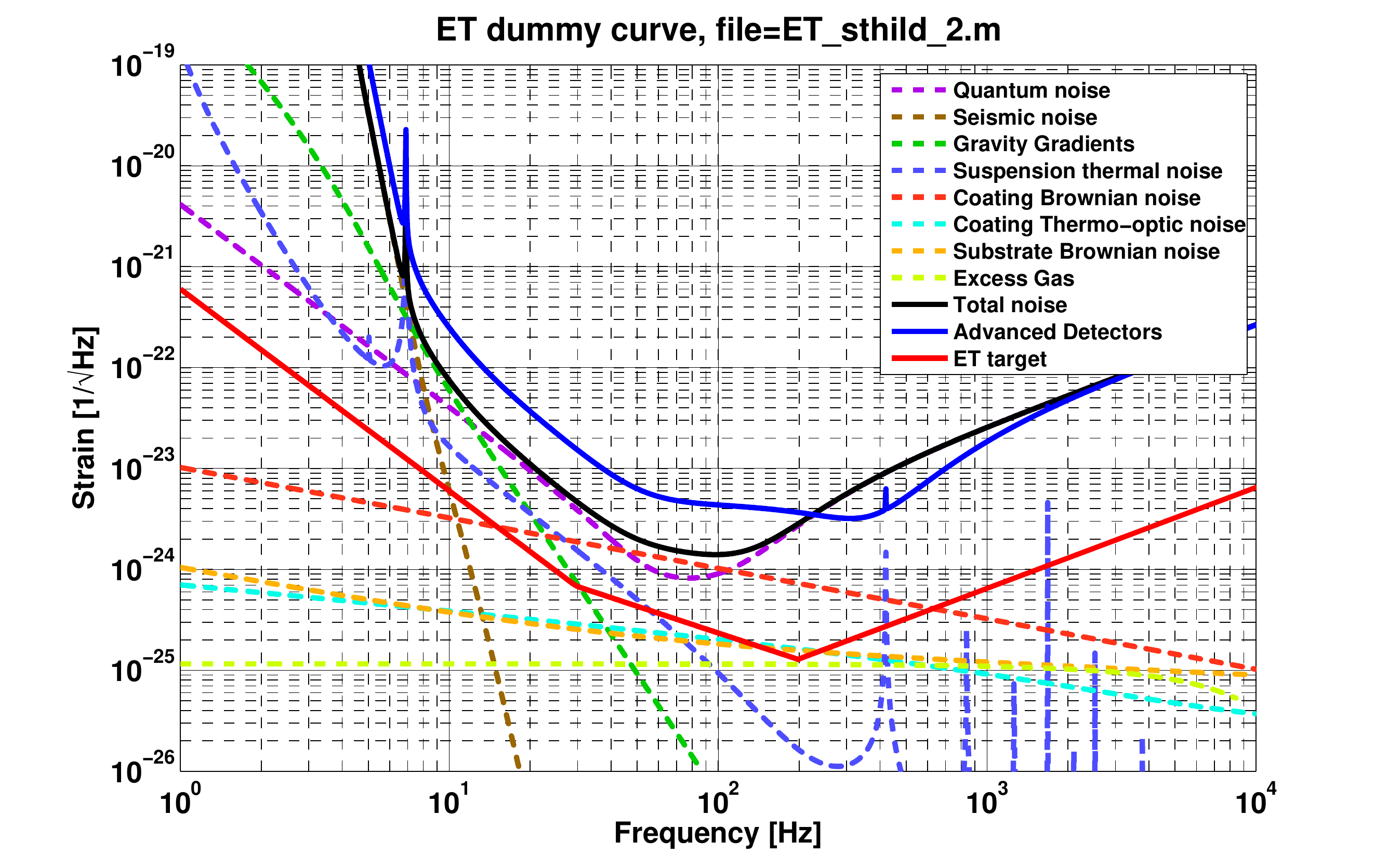}
\includegraphics [viewport=30 0 710 480,scale=0.47]{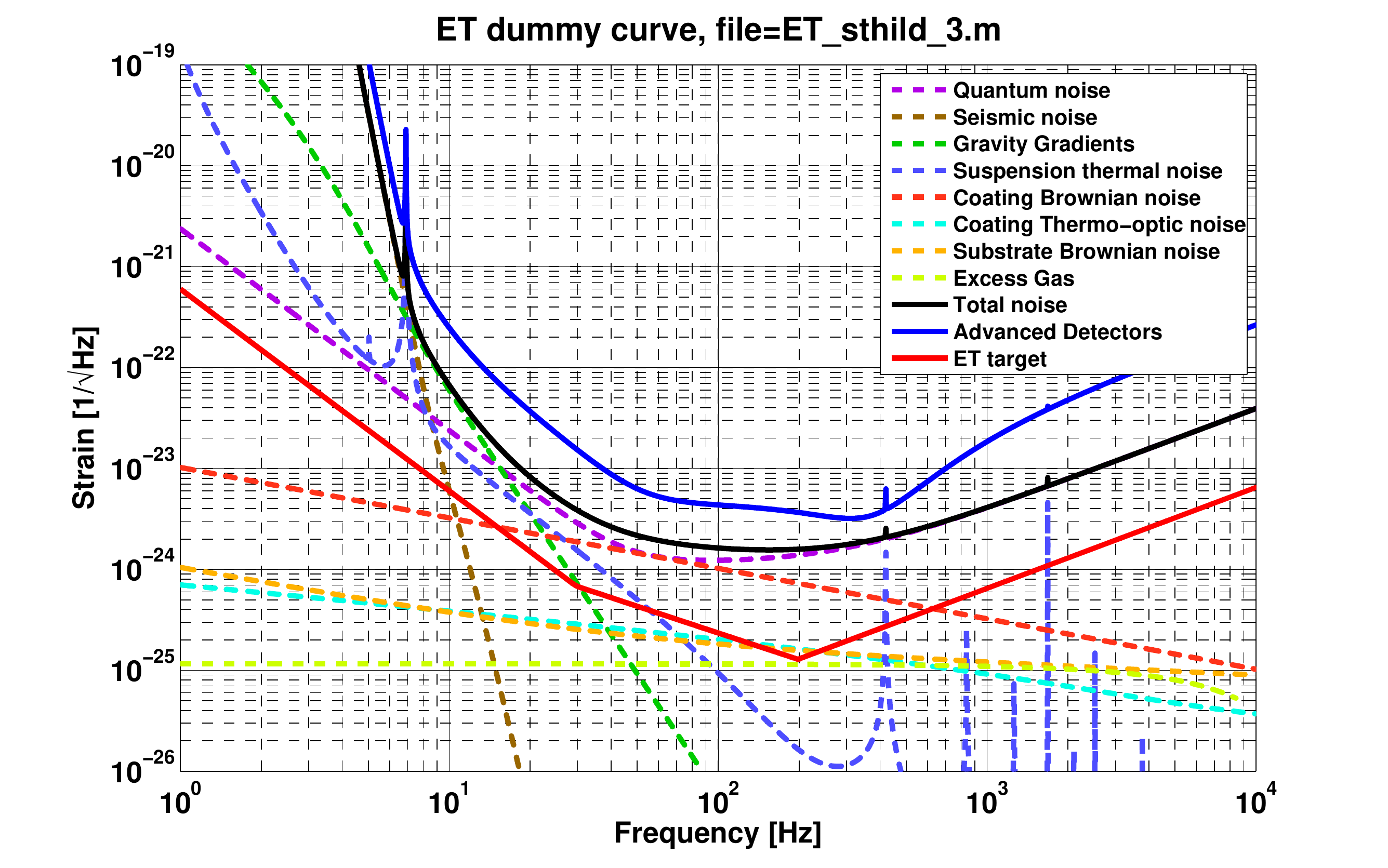}
\end{center}

\clearpage
\newpage

\begin{minipage}{\textwidth}
\includegraphics [viewport=30 0 710 480, scale=0.47,angle=90]{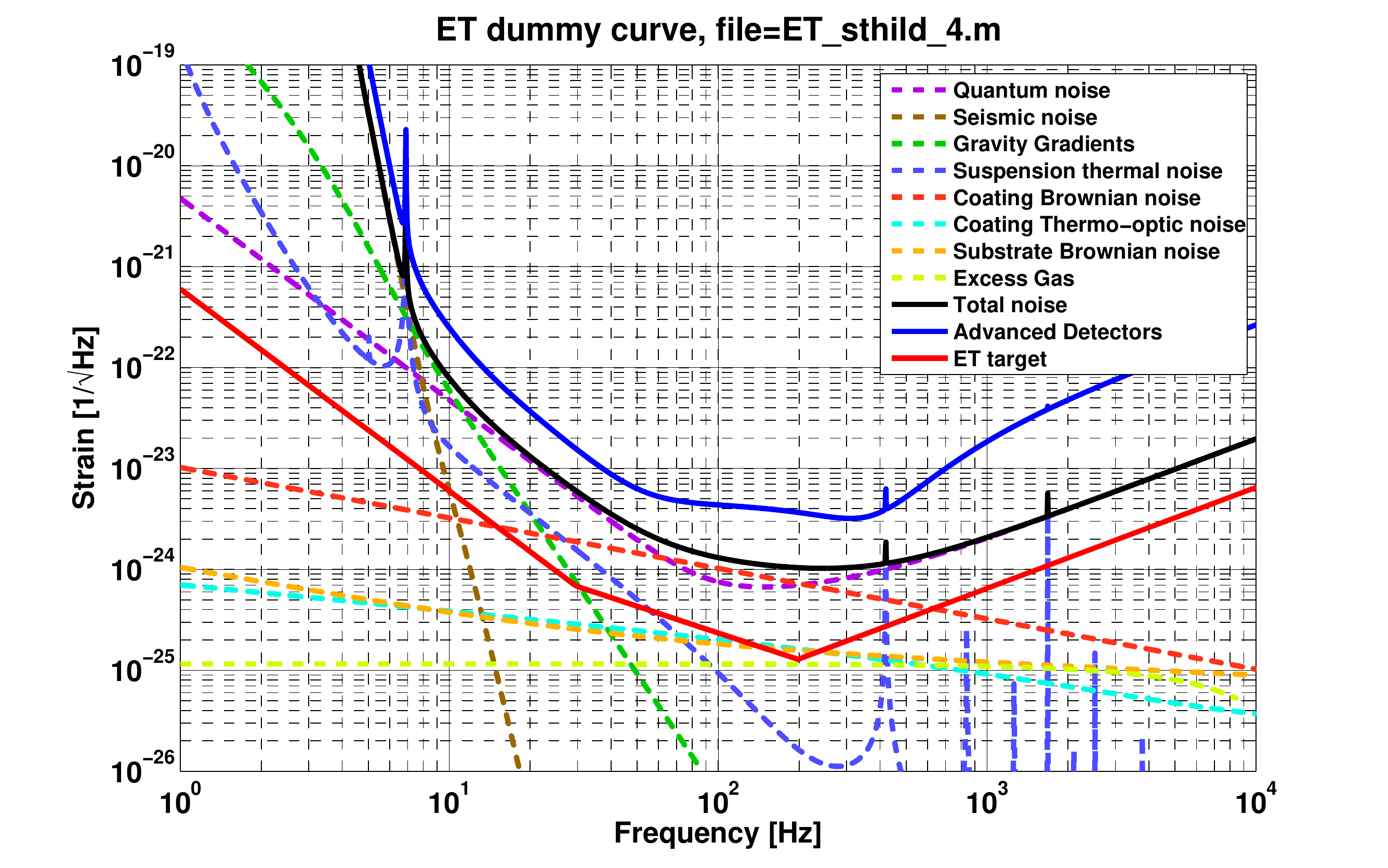}
\includegraphics [viewport=30 0 710 480, scale=0.47,angle=90]{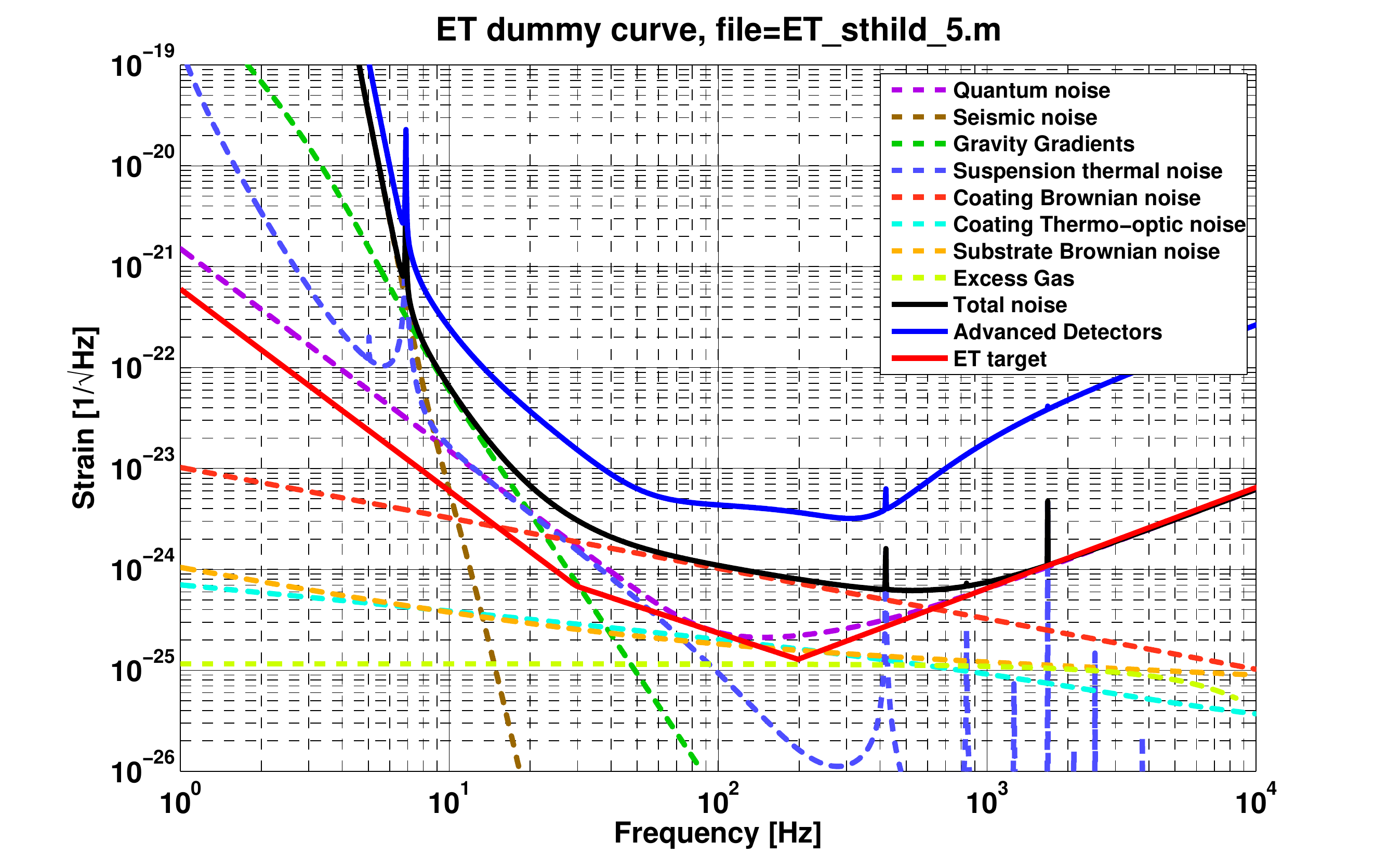}
\end{minipage}

\begin{minipage}{\textwidth}
\includegraphics [viewport=30 0 710 480, scale=0.47,angle=90]{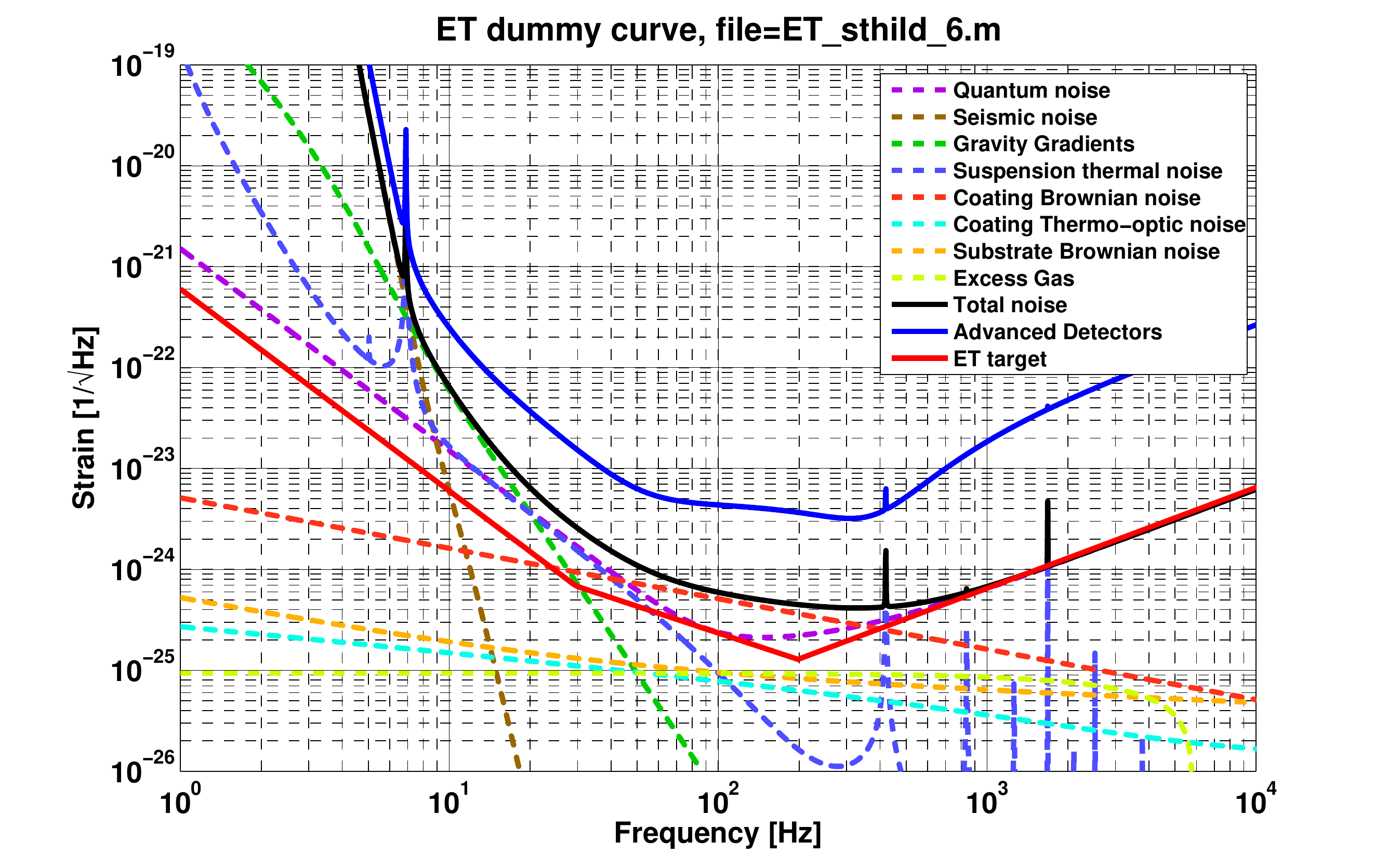}
\includegraphics [viewport=30 0 710 480, scale=0.47,angle=90]{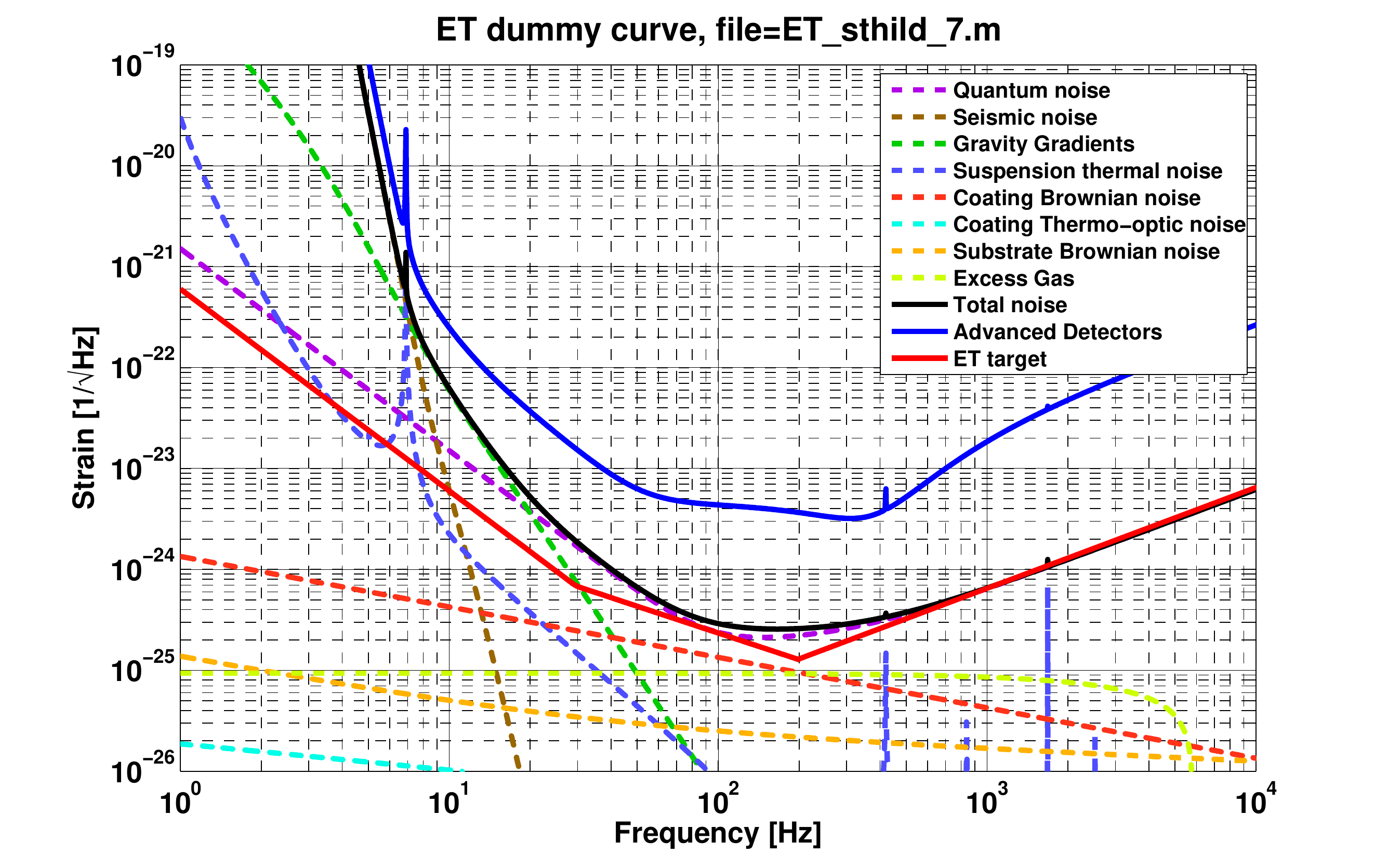}
\end{minipage}
\newpage

\begin{minipage}{\textwidth}
\includegraphics [viewport=30 0 710 480, scale=0.47,angle=90]{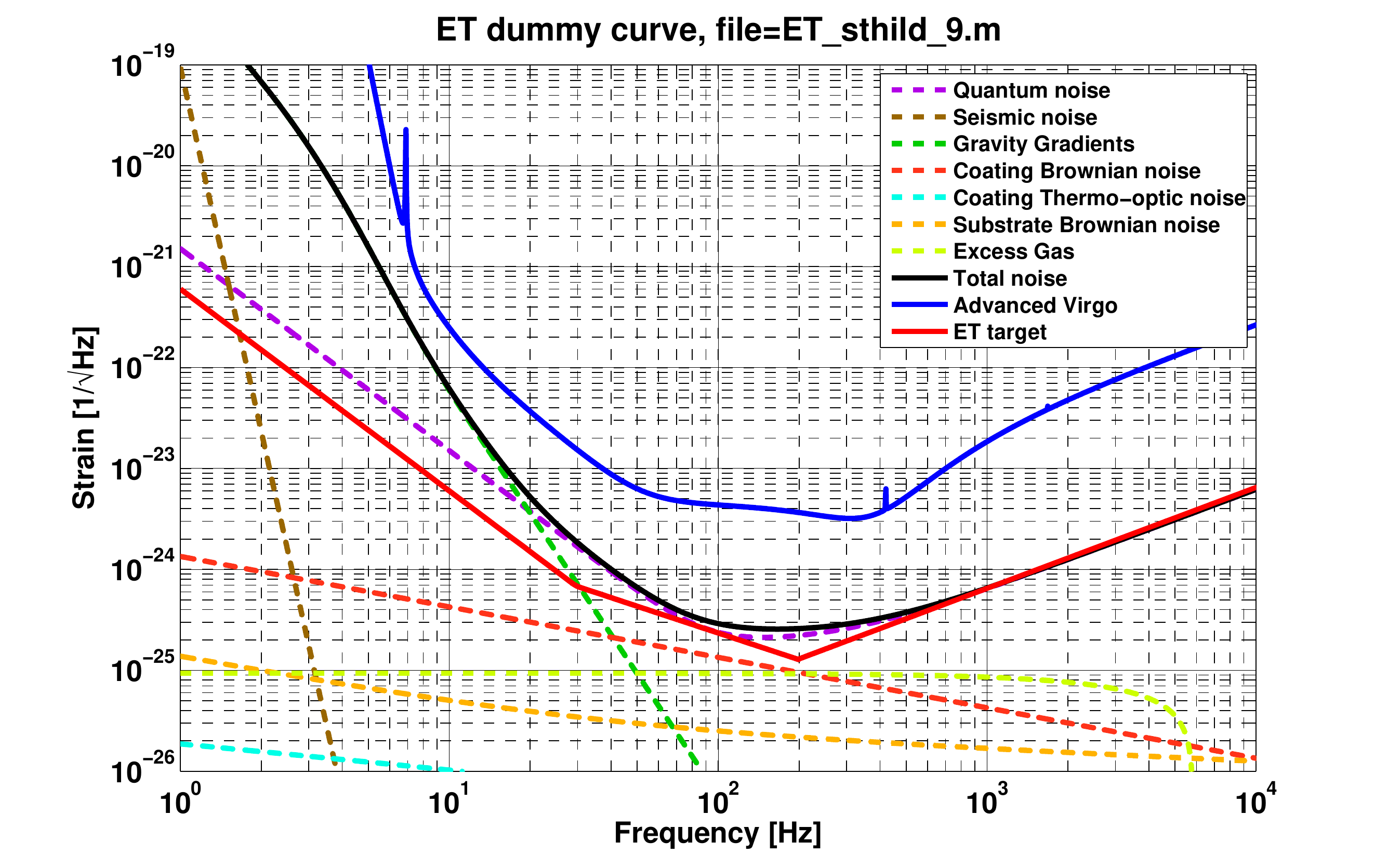}
\includegraphics [viewport=30 0 710 480, scale=0.47,angle=90]{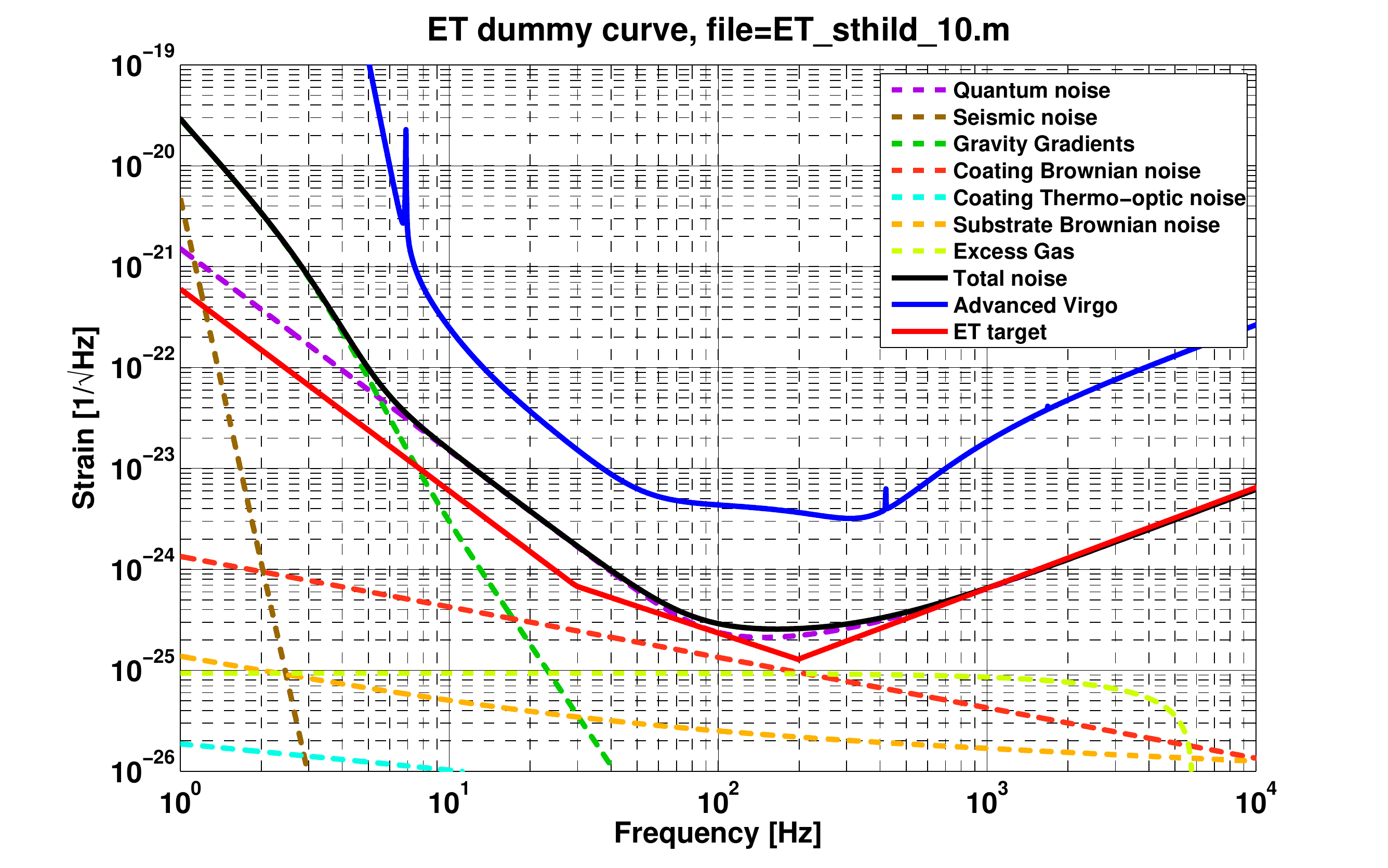}
\end{minipage}

\begin{minipage}{\textwidth}
\includegraphics [viewport=30 0 710 480, scale=0.47,angle=90]{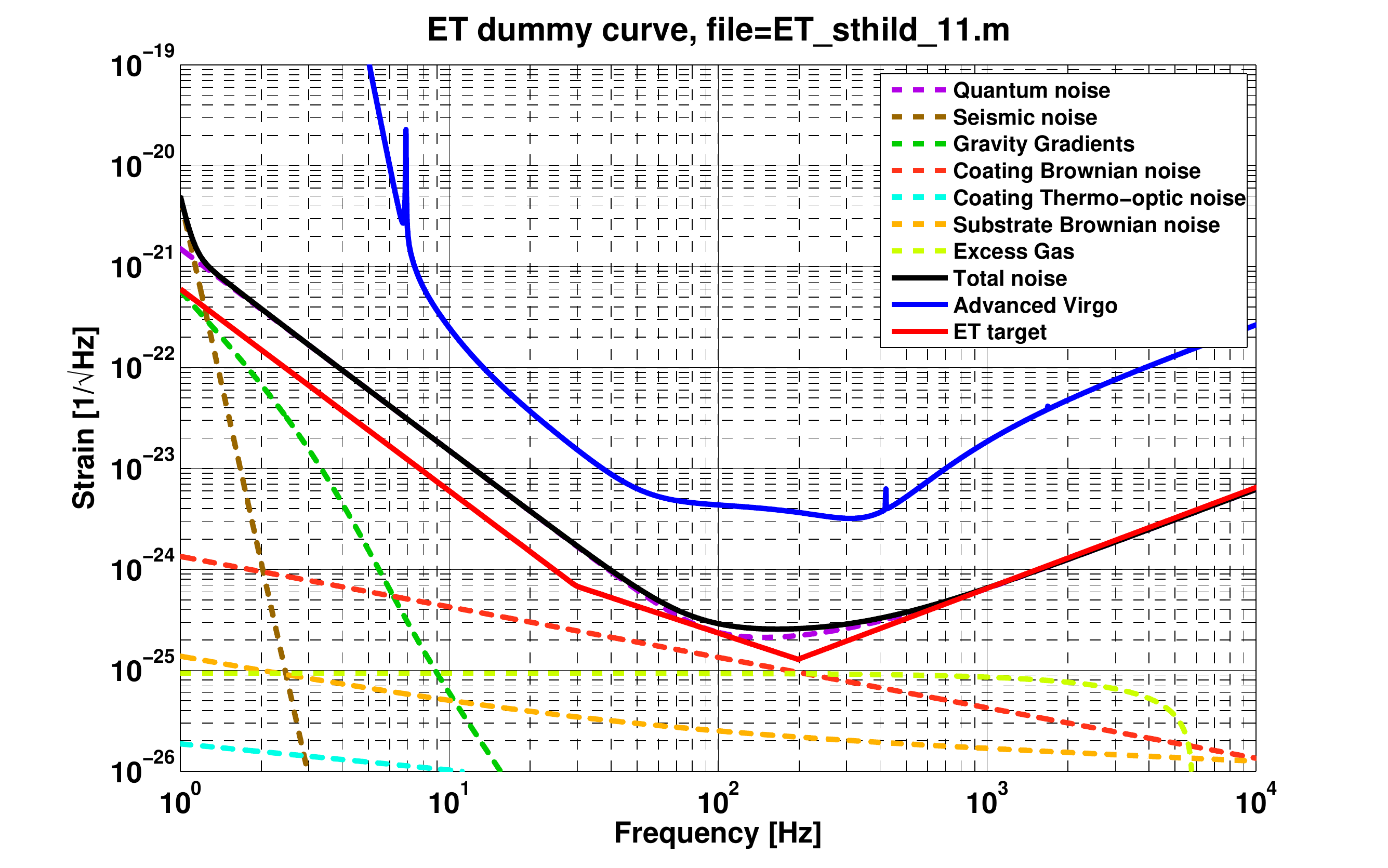}
\end{minipage}

\end{document}